\documentclass[english,reprint,amsmath,amssymb,aps,prb]{revtex4-2}
\usepackage[T1]{fontenc}
\usepackage[utf8]{inputenc}

\usepackage{graphicx}
\usepackage[english]{babel}
\usepackage{xcolor}
\usepackage{siunitx}
\usepackage{physics}

\usepackage[colorlinks=true,citecolor={blue!70!black},linkcolor=black,urlcolor=black]{hyperref}

\begin{document}

\title{The carbon nanotube gatemon qubit}

\author{H. Riechert$^{1}$}
\author{S. Annabi$^{1}$}
\author{A. Peugeot$^{1,2}$}
\author{H. Duprez$^{1}$}
\author{M. Hantute$^{1}$}
\author{K. Watanabe$^{3}$}
\author{T. Taniguchi$^{4}$}
\author{E. Arrighi$^{1}$}
\author{J. Griesmar$^{1}$}
\author{J.-D. Pillet$^{1*}$}
\author{L. Bretheau$^{1}$}

\altaffiliation{These authors jointly supervised this work.
\newline
jean-damien.pillet@polytechnique.edu
\newline
landry.bretheau@polytechnique.edu}

\affiliation{$^{1}$ Laboratoire de Physique de la Mati\`ere condens\'ee, CNRS, Ecole Polytechnique, Institut Polytechnique de Paris, 91120 Palaiseau, France}
\affiliation{$^{2}$ Ecole Normale Sup\'erieure de Lyon, CNRS, Laboratoire de Physique, F-69342 Lyon, France}
\affiliation{$^{3}$ Research Center for Electronic and Optical Materials, National Institute for Materials Science, 1-1 Namiki, Tsukuba 305-0044, Japan}
\affiliation{$^{4}$ Research Center for Materials Nanoarchitectonics, National Institute for Materials Science,  1-1 Namiki, Tsukuba 305-0044, Japan}

\begin{abstract}
Gate-tunable transmon qubits are based on quantum conductors used as weak links within hybrid Josephson junctions. These gatemons have been implemented in just a handful of systems, all relying on extended conductors, namely epitaxial semiconductors or exfoliated graphene.
Here we present the coherent control of a gatemon based on a single molecule, a one-dimensional carbon nanotube, which is integrated into a circuit quantum electrodynamics architecture. The measured qubit spectrum can be tuned with a gate voltage and reflects the quantum dot behaviour of the nanotube. Our ultraclean integration, using a hexagonal boron nitride substrate, results in record coherence times of 200\,ns for carbon nanotube-based qubits.
Furthermore, we investigate its decoherence mechanisms, thus revealing a strong gate dependence and identifying charge noise as a limiting factor.
On top of positioning carbon nanotubes as contenders for future quantum technologies, our work paves the way for studying microscopic fermionic processes in low-dimensional quantum conductors.
\end{abstract}

\maketitle

Superconducting qubits, which have emerged as a leading approach in quantum information science, are based on Josephson tunnel junctions that act as quintessential nonlinear, non-dissipative elements~\cite{Kjaergaard2020}. It is possible to replace the tunnel junction with a hybrid Josephson junction utilizing a quantum conductor as a weak link~\cite{Likharev1979,Tinkham1996}. This approach, backed by recent advances in low-dimensional materials, paved the way for innovative qubit designs like transmon, fluxonium and Andreev qubits~\cite{Larsen2015, DeLange2015, Casparis2016, Kringhøj2018, Luthi2018, Casparis2018, Wang2019,Sagi2024, Kiyooka2025, Hays2018, Tosi2019, Larsen2020, Pita-Vidal2020, Bargerbos2022, Pita-Vidal2023, Bargerbos2023, Strickland2025}. Among these emerging architectures, gatemon qubits stand out~\cite{Larsen2015, DeLange2015, Casparis2016, Kringhøj2018, Luthi2018, Casparis2018, Wang2019, Sagi2024, Kiyooka2025}, offering remarkable functionalities such as qubit frequency tunability through applied gate voltage—providing a practical alternative to magnetic flux control—and enhanced resilience to magnetic fields~\cite{Kroll2018, Kringhøj2021}, making them well-suited for applications in electron and nuclear spin resonance.
Implementing such hybrid circuit quantum electrodynamics (cQED) architectures~\cite{Kurizki2015, Clerk2020}, which combine low-dimensional materials with superconducting circuits, unlocks exciting possibilities at the intersection of quantum information science and condensed matter physics.
The unique electronic properties of these quantum materials indeed shape the qubit's operational behavior, exemplified by the physics of Dirac fermions that is observed in graphene gatemons~\cite{Wang2019}. This synergy also establishes a sensitive platform for probing the microscopic behavior of fermions in quantum materials, relying on minimally invasive microwave signals rather than traditional transport measurements.

To date, the quantum control of gatemon qubits has been demonstrated in devices based on extended conductors, such as semiconducting nanowires, two-dimensional electron/hole gases and graphene~\cite{Larsen2015, DeLange2015, Casparis2016, Kringhøj2018, Luthi2018, Casparis2018, Wang2019, Sagi2024, Kiyooka2025}. Here we propose a gatemon based on a single molecule: a carbon nanotube.
This intrinsically one-dimensional object has a uniquely limited number of internal electronic degrees of freedom, which could play a key role in designing protected qubits~\cite{Vakhtel2023,Vakhtel2024} and supressing decoherence mechanisms, including quasiparticle poisoning—widely regarded as a major barrier to realizing high-coherence gatemons~\cite{Aumentado2023,Zheng2024}.
More importantly, implementing a gatemon with such an elementary junction of only one conduction channel, makes the underlying fermionic physics of the molecule accessible in a controlled environment. The electron-electron interactions play a crucial role in carbon nanotubes, which opens up great prospects in the context of many-body physics~\cite{Deshpande2008, Pecker2013, Sarkany2017, Shapir2019, Arroyo2020, Zhou2024}.
Carbon nanotubes constitute a compelling platform for charge or spin qubit implementation~\cite{Churchill2009, Delbecq2011, Laird2013, Delbecq2013, Viennot2014, Viennot2015, Pei2017, Penfold-Fitch2017, Khivrich2020}. These non-superconducting architectures suffer from short coherence times, $T_2^*$ being typically of the order of 10\,ns, and consequently difficulties in achieving coherent control.
Implementing a superconducting qubit based on a carbon nanotube is thus a promising direction to explore. This requires to integrate a carbon nanotube-based Josephson junction within a cQED architecture, which was realized in a single experimental study so far~\cite{Mergenthaler2021}.
In this work, we report the first coherent control of a nanotube-based gatemon, achieving a record coherence time $T_2^*$ of 200\,ns with significant potential for improvement, as we observe that $T_2^*$ increases exponentially with qubit frequency.
This landmark achievement is attributed to the ultraclean integration of the nanotube within our superconducting circuits, effectively minimizing disorder in the surroundings of the nanotube. To do so, we utilize hexagonal boron nitride (hBN) as a substrate, drawing from advancements in graphene physics~\cite{Dean2010,Wang2019,Wang2022}. This crystalline hBN-nanotube stack is free of atomic-scale defects, which are known to undermine the performance of superconducting qubits~\cite{Lisenfeld2019}. These advancements position carbon nanotube-based gatemons as a compelling platform for exploring and overcoming coherence challenges in next-generation quantum devices.

\section*{Architecture of the circuit}

\begin{figure}
    \includegraphics[width=\columnwidth]{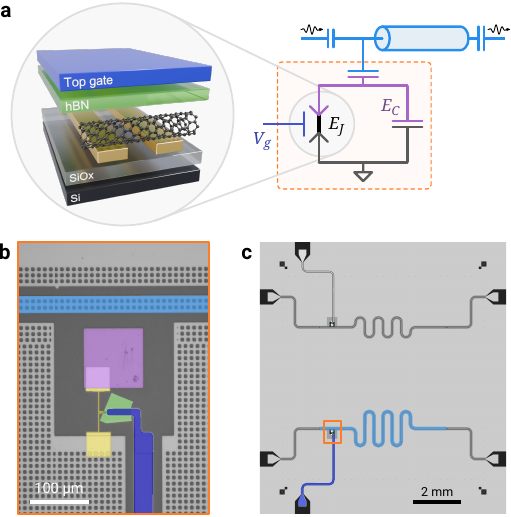}
    \caption{{\bf Fabrication of nanotube gatemon qubits.}
    (a)~Schematic of the hybrid cQED architecture containing a carbon nanotube Josephson junction. The nanotube is transferred onto two Nb\,--\,Au electrodes (gold) to form a Josephson junction of energy $E_J$ that can be tuned with gate voltage $V_g$. Together with the shunt capacitor of energy $E_C$ it implements a gatemon qubit (orange box), which is capacitively coupled to the readout resonator (light blue).
    (b)~False-colored micrograph of the nanotube gatemon. The central charge island (purple), which implements the shunt capacitor, is coupled to a $\lambda/2$ resonator and connected to the ground plane through the carbon nanotube Josephson junction. The carbon nanotube (not visible) is covered by hBN (green) and a top gate (dark blue).
    (c)~Layout of the full chip with two independent hybrid cQED architectures.
    Device~A corresponds to the bottom one, its resonator being highlighted in blue.
    Bonding pad on the bottom of the chip is used for top gate control.
    }
    \label{fig1}
\end{figure}

Our circuit, which is schematized in Fig.~\ref{fig1}a, is primarily composed of two key elements: the nanotube gatemon that functions as a qubit, and a superconducting microwave resonator to readout the qubit state.
The core component of this hybrid cQED architecture is the nanotube Josephson junction.
It is fabricated using a novel technique~\cite{Annabi2024} that integrates an ultraclean single-walled carbon nanotube into prefabricated superconducting circuits on an insulating silicon substrate.
To achieve this, the nanotube, adhered to a thin hBN layer, is mechanically transferred onto two superconducting Nb\,(35\,nm)--Au\,(10\,nm) bilayer electrodes, such that it is encapsulated between the hBN and the silicon substrate.
This configuration not only shields the device from degradation due to air exposure, but also suspends the nanotube between the superconducting electrodes, ensuring that the section carrying the supercurrent avoids direct contact with the substrate.
This arrangement significantly enhances the cleanliness of the device and reduces potential sources of disorder compared to previous works~\cite{Mergenthaler2021}.
The carbon nanotube is capacitively coupled to an aluminum top electrode through the hBN layer, just tens of nanometers thick. The voltage $V_g$ of this top gate controls the chemical potential and, consequently, the electronic density of states in the nanotube. By performing independent transport measurements we have shown that Josephson junctions fabricated using this technique exhibit gate-tunable critical currents \(I_c\), with supercurrents as high as \SI{8}{nA}~\cite{Annabi2024}, comparable to that of conventional transmon qubits.
Nanotube Josephson junctions thus behave as supercurrent field-effect transistors~\cite{Jarillo-Herrero2006}. In the microwave domain, this translates into a gate-dependent non-linear inductance $L_J$ or equivalently into a Josephson energy $E_J=\varphi_0^2/L_J$, where $\varphi_0$ is the reduced flux quantum~\cite{Kringhøj2018}.

The nanotube Josephson junction is embedded into a superconducting circuit, thus forming a hybrid cQED architecture.
In practice, the junction connects between a metallic island and the ground plane of the circuit, as shown in the optical micrograph of Fig.~\ref{fig1}b. This island implements a shunting capacitor of energy $E_C$ and forms, together with the nanotube Josephson junction, an anharmonic oscillator. Its ground and first excited states, $\ket{g}$ and $\ket{e}$, serve as the qubit states at the frequency $f_q\approx \sqrt{8E_JE_C}/h$~\cite{Koch2007}.
For state readout the qubit is capacitively coupled to a coplanar $\lambda/2$ cavity (highlighted light blue in Fig.~\ref{fig1}d). Due to hybridization of the cavity and qubit modes, the cavity's resonance frequency depends on the qubit state and frequency~\cite{Koch2007}, which in turn is tunable by the gate voltage via~$L_J$.
In the following, measurements from two different devices A and B are presented. Device B is similar to device A (shown in Fig.~\ref{fig1}b-c), with a slightly higher cavity resonance, larger qubit-cavity coupling, and an on-chip band-pass filter on the gate line as detailed in Supplementary Section~\ref{app:gate-filter}.

\begin{figure*}
	\includegraphics[width=\textwidth]{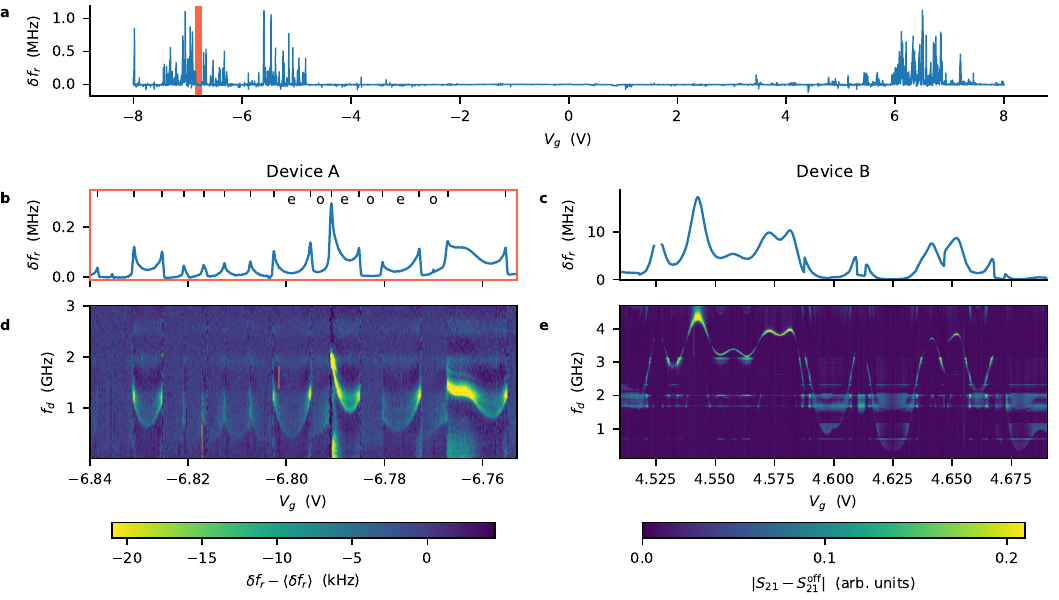}
	\caption{\textbf{Spectroscopy of two gatemon devices.}
    (a)~One-tone spectroscopy of device~A measuring the shift $\delta f_r$ of the cavity at \SI{5.056}{GHz}. Outside a band gap of about \SI{9}{V}, dense forests of positive shifts indicate the presence of the qubit mode at lower frequency. 
    Larger $\delta f_r$ is caused by smaller qubit-cavity detuning due to higher qubit frequency, which stems from larger supercurrent in the carbon nanotube Josephson junction.
    (b,c) One-tone spectroscopy for device~A (close up of (a)) and device B.
    (d,e)~Two-tone spectroscopy of the gate-dependent qubit mode for the same gate voltage ranges. Shown is $\delta f_r$ compared to the vertical average (device~A) or change in cavity transmission $S_{21}$ compared to no second tone drive $S_{21}^\mathrm{off}$ (device~B).
    Spectra of device~A exhibit regularly spaced peaks owing to quantum phase transitions changing the charge parity (letters e/o for even/odd in (b)) of the quantum dot.
    Contrarily, the gate dependence for device~B is smoother, the nanotube junction being in a regime of a superconducting quantum dot strongly coupled to the electrodes.
    Both devices show spurious modes, visible as horizontal lines. The qubit linewidth for frequencies below 2\,GHz is up to $\sim1$\,GHz due to charge dispersion.
    }
	\label{fig:spectroscopy}
\end{figure*}

\section*{Spectroscopy of nanotube gatemon devices}

We first focus on device A and measure the transmission of the $\lambda/2$ cavity in order to extract its resonance frequency $f_r$. Varying the gate voltage $V_g$ modulates the nanotube Josephson junction's inductance \(L_J\), which causes a frequency shift $\delta f_r$ of the resonator.
Fig.~\ref{fig:spectroscopy}a shows this frequency shift $\delta f_r = f_r - f_{r0}$ over a large range of gate voltage, where $f_{r0}=\SI{5.056}{GHz}$ is the bare-resonator frequency. Around $V_g=\SI{0}{V}$, we observe a broad region where \(f_r\) remains largely independent of \(V_g\), which results from the semiconducting gap of the nanotube where the supercurrent is zero and $L_J$ diverges. Beyond this gap, the resonator exhibits positive frequency shifts up to 1\,MHz that varies strongly with $V_g$, as can be seen in the close-up view of Fig.~\ref{fig:spectroscopy}b. This gate-dependent shift indicates a hybridization between the resonator and the nanotube qubit. Fig.~\ref{fig:spectroscopy}c shows the same measurement for device B, with a gate-dependent frequency shift of the resonator that exceeds 10\,MHz.

The cQED architecture provides a powerful platform for probing the nanotube qubit as the readout resonator experiences a frequency shift that depends on the qubit’s state. We thus apply an additional drive signal at frequency \(f_d\) in order to excite qubit transitions from state \(|g\rangle\) to \(|e\rangle\), while monitoring the resonator close to its resonance frequency. Such a two-tone spectroscopy measurement allows for a precise determination of the qubit frequency \(f_q\). Figure~\ref{fig:spectroscopy}d-e show the qubit spectra for device A and B as a function of gate voltage.
They both exhibit a gate-dependence $f_q(V_g)$ that is directly reminiscent from the one observed in cavity spectroscopy (Fig.~\ref{fig:spectroscopy}b-c), which further demonstrates the qubit-resonator hybridization. These spectra allow us to extract for device A (respectively B) a charging energy $E_C=\SI{260}{MHz}$ (resp. 330\,MHz) and a qubit-cavity coupling strength $g\approx \SI{50}{MHz}$ (resp. 120\,MHz)~\cite{Koch2007}. 
The stronger coupling and higher qubit frequency of device B explain the greater $\delta f_r$ measured in Fig.~\ref{fig:spectroscopy}c.
Crucially, both spectra display a large tunability of the qubit frequency \(f_q\) over 4\,GHz, demonstrating that the qubit behaves as a gatemon.
This tunability arises from the fact that the nanotube Josephson energy $E_J$ depends on the gate voltage, and can be varied between a few hundred of MHz and up to \SI{8}{GHz}.
Using the simple relation $E_J=\varphi_0 I_c$ (valid for a tunnel junction), the critical current of the nanotube junctions is estimated to be $\sim 0.6-16\,\text{nA}$. These values are consistent with results reported in transport measurements~\cite{Annabi2024}, taking into account the known discrepancy between switching current and critical current in small junctions. 

More fundamentally, the voltage-tunability of $E_J$ originates from the electrostatic control of the Andreev bound states, which are localized in the nanotube and are responsible for the Josephson effect~\cite{Pillet2010}. This fermionic origin is revealed by analyzing the gate-dependent spectra in Fig.~\ref{fig:spectroscopy}, which are very different for the two devices.
For device A, the spectra exhibit sharp peaks when varying $V_g$, especially visible in Fig.~\ref{fig:spectroscopy}b.
These are a hallmark of quantum dot behavior, where the nanotube exhibits characteristics akin to Coulomb blockade effect commonly seen in transport measurements. Each peak indeed corresponds to the addition of a single charge to the nanotube, when an electronic level within the quantum dot is brought into resonance with the Fermi energy of the electrodes. The abrupt frequency jumps at the peaks are reminiscent of $0-\pi$ quantum phase transitions observed in supercurrent measurements of nanotube junctions~\cite{vanDam2006, Jrgensen2007, Maurand2012, Delagrange2016}. Such a phenomenon is predicted to occur in interacting quantum dot-based Josephson junctions when the coupling~$\Gamma$ of the dot to the superconducting electrodes is moderate compared to the Coulomb repulsion~$U$ \cite{Vecino2003, Meng2009, Zazunov2010}.
The competition between pairing and repulsion interaction results in sudden changes of the fermionic ground state with different parities.
Even parity Andreev bound states carry more supercurrent than odd parity ones, which translates into jumps of~$f_q$. This allows us to assign to each gate voltage region a parity for the charge state of the carbon nanotube, as shown in Fig.~\ref{fig:spectroscopy}b.
For device B (Fig.~\ref{fig:spectroscopy}c,e), the spectra exhibit a smoother gate-dependence. In that case, almost no parity-changes are observed and the nanotube behaves as a weakly-interacting quantum dot, whose electronic levels are tuned by $V_g$ causing smooth oscillations in $f_q$.
This suggests that the ratio $\Gamma/U$ is here larger compared to device~A, which we interpret as a stronger coupling~$\Gamma$ to the electrodes. The latter translates into an increased Josephson energy $E_J$, thus explaining the higher observed qubit frequency which reaches up to 4.3\,GHz.
Finally, the smaller linewidth observed at higher qubit frequency promises better coherence properties.

\begin{figure*}
    \centering
    \includegraphics[width=\textwidth]{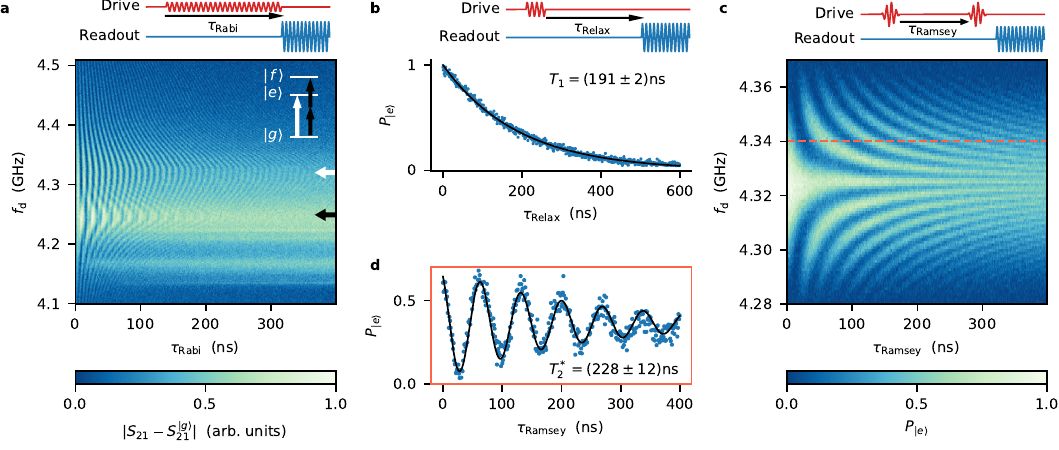}
    \caption{{\bf Quantum control of a nanotube gatemon qubit.}
    (a)~Rabi oscillations as a function of drive frequency $f_d$ and pulse duration \(\tau_{\text{Rabi}}\). The two chevrons pattern observed correspond to qubit transition at \SI{4.32}{GHz} (white arrow) and two-photon transition to the second excited state \SI{71}{MHz} below (black arrow).
    The signal is the change of cavity transmission $S_{21}$ compared to the ground state transmission $S_{21}^{\ket{g}}$.
    (b)~Relaxation measurement (blue points) following a $\pi$-pulse. The exponential fit (black line) leads to $T_1=191\pm2\,\text{ns}$.
    (c)~Ramsey oscillations as a function of drive frequency $f_d$ and time delay \(\tau_{\text{Ramsey}}\) between two $\pi/2$ Gaussian pulses. The oscillation frequency is given by the detuning of the drive from the qubit frequency.
    (d)~Ramsey oscillations measured at $+\SI{20}{MHz}$ detuning (blue points), which is a line cut in (c) indicated as dashed line. The fit of an exponentially decaying cosine (black line) results in a coherence time of $T_2^*=228\pm12\,\text{ns}$.
    All measurements were performed on device~B at $V_g=\SI{-4.2376}{V}$.
    }
    \label{fig:coherent-control}
\end{figure*}

\section*{Quantum control and coherence measurements}

To demonstrate that a system functions as a two-level system or qubit, it is essential to establish its coherent control in the time domain. We focus in the following on device B that has shown the best coherence, and first investigate it at a constant gate voltage. Similar measurements for device~A are available in Supplementary Section~\ref{app:coherence-deviceA}.
We drive the qubit with an initial pulse at frequency \(f_d \sim f_q\) for a duration \(\tau_{\text{Rabi}}\), followed by a second pulse at the resonator frequency \(f_r\) for qubit readout.
Figure~\ref{fig:coherent-control}a, which shows the qubit state as a function of both \(\tau_{\text{Rabi}}\) and $f_d$, exhibits the characteristic chevron pattern associated with Rabi oscillations. At resonance $f_d=f_q$ (white arrow at \SI{4.32}{GHz}), the qubit state undergoes coherent oscillations in the Bloch sphere from \(|g\rangle\) to \(|e\rangle\) at the Rabi frequency $\Omega_R=\SI{107}{MHz}$. Driving with a detuning \(\delta = f_d - f_q\) induces oscillations at higher frequency $\Omega = \sqrt{\Omega_R^2 + \delta^2}$ with a reduced contrast.
The measurement reveals a second chevron pattern centered at \(f_{gf}/2\approx \SI{4.25}{GHz}\) corresponding to a two-photon transition of energy \(hf_{gf}\) between the ground state \(|g\rangle\) and the second excited state \(|f\rangle\).
The two-photon nature of the transition is evident in the enhanced $\delta$ dependence of the Rabi frequency, $\Omega = \sqrt{\Omega_{gf}^2 + 4\delta_{gf}^2}$, where $\Omega_{gf} \approx \SI{55}{MHz}$ denotes the two-photon Rabi frequency at resonance and $\delta_{gf}=f_d-f_{gf}/2$~\cite{Linskens1996}.
From this measurement, we can extract a qubit anharmonicity $\alpha=f_{gf}-2f_q$ of \SI{-142}{MHz}. We have measured values of $\alpha$ between \SI{-62}{MHz} and \SI{-201}{MHz} for various gate voltages (see Supplementary Section~\ref{app:anharmonicity}). This metric, which was too small to be extracted in the case of the graphene gatemon~\cite{Wang2019}, is critical to evaluate a qubit quality as it determines a lower bound for a pulse duration. Contrary to the case of the tunnel junction-based transmon, the anharmonicity $\alpha$ is not simply given by $-E_C$ but varies with $V_g$ as it depends on microscopic parameters of the nanotube gatemon~\cite{Kringhøj2018}.
Interestingly, we observe a large deviation of $|\alpha|$ with respect to $E_C$, with values that can go below the standard boundary $E_C/4$~\cite{Kringhøj2018}. This uncommon observation could be related to the quantum dot nature of our carbon nanotube Josephson junction, as recently predicted in Ref.~\cite{Fatemi2024}.

Going further, we investigate the coherence properties of the nanotube gatemon.
The Rabi oscillations are used to calibrate the $\pi$ and $\pi/2$-pulses, which allow us to prepare the qubit in the states $\ket{e}$ and $\left(\ket{g}+\ket{e}\right)/\sqrt2$ respectively.
We first apply a $\pi$-pulse followed by a readout pulse delayed by the time $\tau_{\text{Relax}}$, which results in an exponential decay of the qubit state as shown in Fig.~\ref{fig:coherent-control}b. The extracted $T_1=\SI{191}{ns}$ indicates the characteristic duration for the qubit to lose energy and relax to its ground state.
Next, we measure the qubit dephasing using Ramsey interferometry, where we apply two $\pi/2$-pulses at frequency $f_d$ separated by a waiting time $\tau_{\text{Ramsey}}$. Figure~\ref{fig:coherent-control}c-d display Ramsey oscillations at the detuning frequency $\delta = f_d - f_q$, which correspond to precession of the qubit state in the Bloch sphere at the equator. From their decay envelope we extract a coherence time $T_2^*=\SI{228}{ns}$.
The similar timescale of $T_1$ and $T_2^*$ indicates that decoherence originates both from energy relaxation and pure dephasing.

\begin{figure}
    \centering
    \includegraphics[width=\columnwidth]{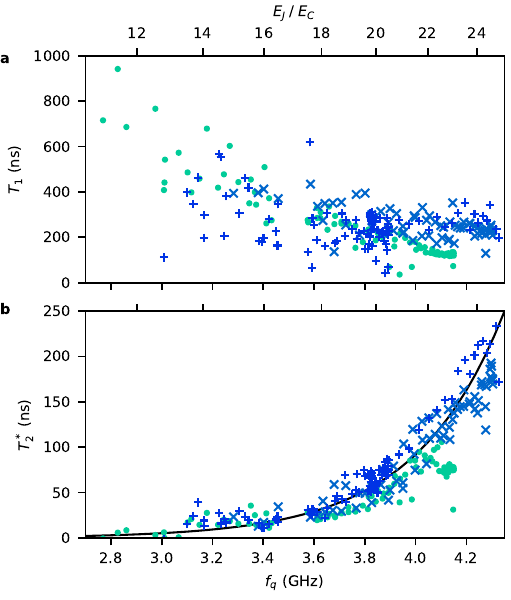}
    \caption{{\bf Coherence time measurements of a nanotube gatemon at many gate voltages.}
    (a)~Relaxation time $T_1$ for three different gate regions, shown by different markers (see associated qubit spectroscopy in Supplementary Section~\ref{app:spectro-deviceB}).
    (b)~Coherence time $T_2^*$ corresponding to the same gate points. Both times $T_1$ and $T_2^*$ are plotted as a function of qubit frequency $f_q$ (bottom horizontal axis) and corresponding $E_J / E_C$ (top horizontal axis), using $h f_q = \sqrt{8E_J E_C} - E_C$. The exponential behavior of $T_2^*\propto e^{hf_q/E_C}$ (black line) is due to charge dispersion. All measurements were performed on device~B.
    }
    \label{fig:gate-dependence}
\end{figure}

To gain deeper insight into the mechanisms limiting coherence, we extensively measured our nanotube gatemon at many values of gate voltage~$V_g$. Figure~\ref{fig:gate-dependence} shows the relaxation time $T_1$ and the coherence time $T_2^*$ acquired for a large set of $V_g$ and plotted as a function of the qubit frequency~$f_q$.
The relaxation time measurement exhibits a global trend, with $T_1$ increasing at low qubit frequency and reaching up to \SI{942}{ns}. The origin of this dependence is not known. More importantly, Fig.~\ref{fig:gate-dependence}a shows a large spread of~$T_1$ values, with no strong correlation in~$f_q$. Energy relaxation is here likely limited by microscopic mechanisms that are gate-dependent.
On the contrary, the coherence time measurements exhibit a strong correlation, as $T_2^*$ increases with $f_q$ and reaches up to \SI{233}{ns}. We can explain this dependence using the transmon model~\cite{Koch2007} that predicts a dephasing time scaling as $e^{\sqrt{8E_J/E_C}}$. The range of $f_q$ spanned in Fig.~\ref{fig:gate-dependence}b corresponds to a ratio $E_J/E_C$ varied between 11 and~25. We thus model our data by $T_2^*\propto e^{hf_q/E_C}$ (black curve in Fig.~\ref{fig:gate-dependence}b), using the fact that the coherence time is here much lower than $2T_1$ and thus dephasing-limited.
Note how at lower $f_q$ the smaller $E_J/E_C$ ratio also manifests itself in a broad qubit linewidth due to offset charge noise, as evident in Fig.~\ref{fig:spectroscopy}d-e.
We thus demonstrate that our qubit can be gradually tuned from a Cooper pair box to the transmon regime~\cite{Koch2007}.
Further analysis reveals that $T_2^*$ is not limited by gate noise on $E_J$ (see Supplementary Section~\ref{app:gate-noise}). This detailed understanding provides a clear path for improving the nanotube gatemon coherence.

\section*{Conclusions}

In summary, we have demonstrated quantum control of a superconducting qubit made with an ultraclean carbon nanotube-based Josephson junction.
The nanotube-based gatemon qubit exhibits a voltage tunability of its frequency over more than \SI{4}{GHz} and an anharmonicity up to \SI{200}{MHz}.
Time-domain measurements establish quantum coherence of the nanotube gatemon, resulting in a coherence (relaxation) time as high as \SI{233}{ns} (\SI{942}{ns}), an improvement by a factor 4~(27) compared to its graphene counterpart~\cite{Wang2019}.
This suggests that reducing the number of conduction channels effectively suppresses certain decoherence mechanisms, for example the coupling to spurious two-level systems or poisoning of Andreev bound states in the weak link by non-equilibrium quasiparticles~\cite{Zgirski2011,Aumentado2023}.
Moreover, these coherence times approach the first-generation gatemons that were based on semiconducting nanowires~\cite{Larsen2015}. Increasing coherence would require more systematic studies~\cite{Luthi2018,Feldstein-Bofill2024}, a trend that is likely to be followed by nanotube gatemons.
More generally, these results make our device the most coherent carbon-based qubit ever implemented~\cite{Baydin2022} and the first one to be integrated within a cQED architecture, which opens the way towards long-range coupling mediated by microwave photons. This marks a pivotal step towards harnessing single molecules for quantum computing applications.

Going further, our investigations reveal the critical role of charge noise in limiting coherence, suggesting that optimizing the ratio \(E_J/E_C\) could enhance qubit performance. We anticipate substantial improvements in coherence by optimizing microwave engineering of the gatemon environment and by refining nanotube integration in order to minimize disorder. The latter could be achieved by employing higher-quality substrates or implementing bottom gates to shield against defects and stray charges. These efforts will strengthen the nanotube's potential as a promising candidate for innovative quantum technologies. Future progress should be achieved by exploring diverse architectures relying on nanotube Josephson junctions, such as fluxonium or Andreev qubits.

More fundamentally, designing hybrid superconducting circuits that integrate low-dimensional quantum materials represents a promising approach to explore the underlying microscopic fermionic processes at play. We could here reveal the quantum dot behavior of the nanotube in the qubit spectrum.
Interestingly, no coherence of the gatemon qubit could be measured when the parity of the fermionic ground state is odd. On top of the role played by offset charge noise at low $E_J/E_C$, this could originate from the degeneracy of the odd fermionic Andreev ground state.
We plan in the future to probe in a highly sensitive and non-invasive way the Andreev bound states that form in the nanotube~\cite{Pillet2010, Janvier2015, Tosi2019, Hays2020, Pita-Vidal2023}, as well as nonlocal states in Andreev molecules~\cite{Pillet2019, Haxell2023} and more generally the spin and valley degrees of freedom of carbon nanotubes~\cite{Laird2013, Penfold-Fitch2017, Neukelmance2024}. Going further, such platforms should make it possible to study many-body physics that arise from Coulomb repulsion or intriguing topological phases that are predicted to form in one-dimensional materials~\cite{Klinovaja2011, Klinovaja2012, Marganska2018}.

\section*{Data availability}
The data that support the findings of this work and the analysis code to reproduce the figures are available from the Zenodo repository at \url{https://doi.org/10.5281/zenodo.15584225}~\cite{dataset}.

\begin{acknowledgments}
We thank J.\,I.\,J. Wang and W.\,D. Oliver from the EQuS group at MIT with their invaluable help on the microwave design.
Gratitude is extended to S. Cances for participation to the project in its early stages.
We thank D. Markovi\'c, R. Ribeiro-Palau, and the SPEC of CEA-Saclay, in particular the Quantronics group, for their help on nanofabrication and microwave expertise, and to D.~Roux and R.~Mohammedi from LPMC for their technical support.
L.\,B. acknowledges support of the European Research Council (ERC) under the European Union's Horizon 2020 research and innovation programme (Grant Agreement No.~947707). 
J.-D.\,P. acknowledges support of Agence Nationale de la Recherche through grant ANR-20-CE47-0003. 
This work has been supported by the French ANR-22-PETQ-0003 Grant under the France 2030 plan.
K.\,W. and T.\,T. acknowledge support from the JSPS KAKENHI (Grant Numbers 21H05233 and 23H02052), the CREST (JPMJCR24A5), JST and World Premier International Research Center Initiative (WPI), MEXT, Japan.
\end{acknowledgments}

\section*{Supplementary Material}
\begin{appendix}
\renewcommand{\thesection}{\Roman{section}}

\section{Circuit fabrication}
We now describe the fabrication recipe of the devices presented in this paper.

\paragraph{Substrate}
Each circuit is fabricated on a silicon chip cut from an intrinsic silicon wafer with 150\,nm thermally grown oxide.
The wafer is covered by a 155\,nm Nb film deposited via sputtering.
For the dicing into $10.8\times9.8\,\text{mm}^2$ chips, the wafer is protected with a layer of UVIII or S1813 resist, that is subsequently removed with acetone and isopropyl alcohol. This wafer differs from the one used in Ref.~\cite{Annabi2024}, where the bulk was conductive and served as a back gate, but was not well-suited for designing microwave resonators. For this reason we use local top gates in our design instead of a global back gate.

\paragraph{Nb film etching}
The majority of our circuit is patterned into the Nb film using reactive ion etching through a resist mask. This mask is defined by optical laser lithography on a S1813 layer. The development is carried out in MF319 for 60\,seconds and stopped in deionized water. The etching step is then performed using a $\text{CF}_4$/Ar gas flow, ensuring complete removal of the Nb film down to its full thickness. Afterward we remove the resist with N-Methyl-2-pyrrolidone (NMP) at $\SI{70}{\degreeCelsius}$ and then remaining residues with an $\text{O}_2$ plasma ashing process.

The resulting circuit includes the ground plane, the microwave resonator, the feed lines for the top gates, and the superconducting island of the gatemon. The only missing element is the carbon nanotube Josephson junction with its top gate.

\paragraph{Nanotube Josephson junction}
The nanotube Josephson junction is formed, as explained in the main text, by depositing a carbon nanotube onto Nb–Au superconducting electrodes using a hBN flake, following the same method as in Ref.~\cite{Annabi2024}.  

The electrodes are deposited by e-beam evaporation through a bilayer MMA/PMMA mask, defined by electron beam lithography and developed in a 1:3 MIBK:IPA solution.

\paragraph{Top gate}
The top gate was fabricated differently for devices~A and~B.

For device~A, the hBN surface was roughened by a short reactive ion etching with $\text{CHF}_3$ and $\text{O}_2$. Then, a 100 nm\,Al top gate was evaporated after argon ion milling, which ensured a galvanic contact between the top gate and its feed line. In the same step, metal patches were deposited to connect the electrodes to the niobium ground plane and the qubit capacitor.

For device~B, no etching was applied before evaporating a bilayer consisting of an 8\,nm Ti adhesion layer followed by 100\,nm Al. In a separate step, reconnection patches for the top gate and electrodes were defined using argon ion milling and the evaporation of 120 nm\,Al.

\begin{figure}
	\includegraphics[width=\columnwidth]{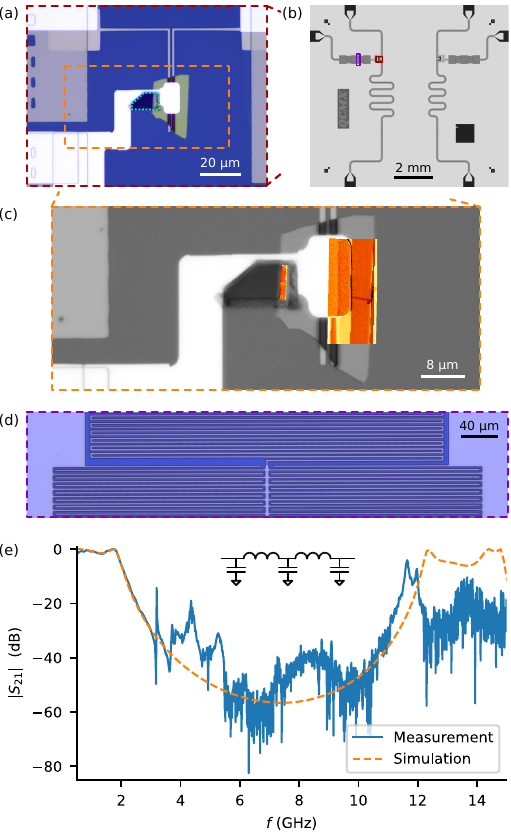}
	\caption{\textbf{Device B.} (a)~Micrograph of the qubit on device B. Part of the top gate needed to be etched (outlined in cyan) to break an accidental contact to the tube protruding from under the hBN (green). (b)~Full device layout. (c)~Atomic/electrostatic force microscopy images overlaid in orange on top of a grayscale micrograph. The carbon nanotube, covered by hBN, appears as dark line  on the right and bright line on the left edge of the top gate. (d)~Micrograph showing a section of the lumped-element gate filter, fabricated with a resolution of \SI{2}{\um}. The upper part is the inductor, the lower part the finger capacitor. (e)~Transmission through the gate filter as simulated and measured in a test device. The inset shows the circuit diagram of the filter.}
	\label{fig:deviceB}
\end{figure}

\paragraph{Removing electrical connection between top gate and nanotube (device B)}
A micrograph of device B is shown in Fig.~\ref{fig:deviceB}. Initially, the top gate was unintentionally electrically connected to one end of the carbon nanotube, which extended a few nanometers beyond the hBN boundary. To break this contact, additional wet etching of the aluminum top gate and ion milling of the titanium adhesion layer were required in the area highlighted in Fig.~\ref{fig:deviceB}a.

The additional fabrication steps performed after nanotube transfer may have introduced greater disorder in the device. This could partly explain why our measurements indicate a lower device quality compared to the transport measurements in Ref.~\cite{Annabi2024} where fourfold degeneracy could be seen.

\section{On-chip filtering on the top gate of device B}
\label{app:gate-filter}

To extend the qubit lifetime in device~B compared to device~A, a band-stop filter is integrated into the gate line of device~B. It is fabricated in niobium alongside the rest of the cQED architecture and consists of a fifth-order lumped-element LC filter, providing at least \SI{-20}{dB} attenuation within a stopband from 1.8\,GHz to 12\,GHz. The wide stopband ensures effective isolation of the tunable qubit across a large portion of its frequency range.

A section of the inductors and capacitors is shown in Fig.~\ref{fig:deviceB}c. Electromagnetic simulations of the structure closely match transmission measurements of the filter on a test device (see Fig.~\ref{fig:deviceB}d).

\begin{figure}
	\includegraphics[width=\columnwidth]{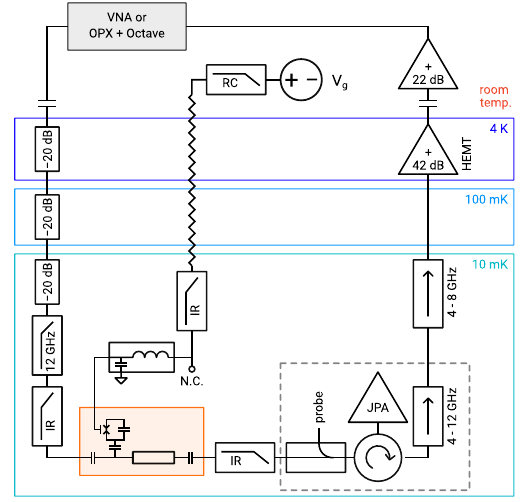}
	\caption{\textbf{Schematic of experimental setup.} The orange box represents the qubit devices discussed in the main text. The Josephson parametric amplifier (dashed box) was only used for measurements of device~B. Straight connections are SMA cables or BNC (gate line at room temperature), the zig-zag line corresponds to a Thermocoax cable. Not shown are additional connections for pumping and biasing the JPA's flux as well as a line to probe the JPA for calibration.}
	\label{fig:cryostat-setup}
\end{figure}

\section{Experimental setup}
The schematic of the setup is shown in Fig.~\ref{fig:cryostat-setup}. Note that the sample is mounted inside a JAWS sample holder~\cite{Villiers2023}.

Spectroscopy measurements of device A are taken with a Rhode \& Schwarz ZNB20 vector network analyzer~(VNA). All other measurements are taken using an \mbox{OPX-1} and Octave from Quantum machine. For measurements of device~B a Josephson parametric amplifier~(JPA) from Quantum microwaves was used. All lines to the sample are filtered with infrared low-pass filters from BlueFors and the input line with an additional 12\,GHz K\,\&\,L low-pass filter.

The gate voltage is applied using the SP927 low-noise voltage source from Basel Instruments. The gate line is filtered at room temperature with a 1.3\,kHz third order RC low-pass filter. Directly connected to the sample holder is a bias-T on the gate-line with a 20\,kHz cutoff frequency. By grounding the high-frequency branch we make a reflective filter that improves the qubit coherence. A SMA tee-connector, which is left open when measuring, allows grounding the gate to protect the sample against electrostatic discharge while working on the setup.

\section{Readout of qubit A and B}
Spectroscopic and time-domain measurements on device A are performed in the dispersive limit, with readout at the resonator frequency $f_r$ and power set low enough to avoid a power-dependent resonance shift in the cavity.

Time-domain measurements for device B are carried out using qubit ``punch-out'' technique: readout at the bare resonator frequency $f_{r0}$ and power optimized to achieve a large signal difference between the driven and non-driven qubit~\cite{Reed2010}.

\begin{figure}
    \centering
    \includegraphics[width=\columnwidth]{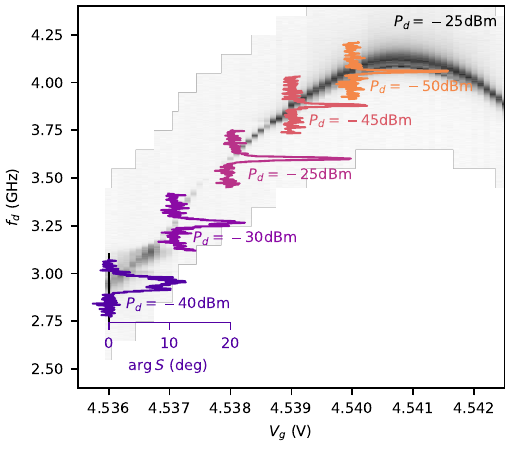}
    \caption{\textbf{Qubit linewidth.} Line cuts of the qubit spectroscopy at increasing qubit frequency show a decreasing qubit linewidth. The gray color scale is $\arg S_{21}$ after removing the vertical average. Because the power delivered to the sample varies strongly with frequency, the drive power $P_d$ is adjusted for each line cut such that $\arg S_{21}$ is significant but small enough to be proportional to the cavity shift.}
    \label{fig:qubit-linewidth}
\end{figure}

\section{Qubit linewidth}
The change of the qubit linewidth versus the qubit frequency is shown in Fig.~\ref{fig:qubit-linewidth}. It is caused by the reducing charge dispersion for larger $f_q$ and thus directly related to the behavior of the coherence time observed in Fig.~4 of the main text.

\begin{figure}
	\includegraphics[width=\columnwidth]{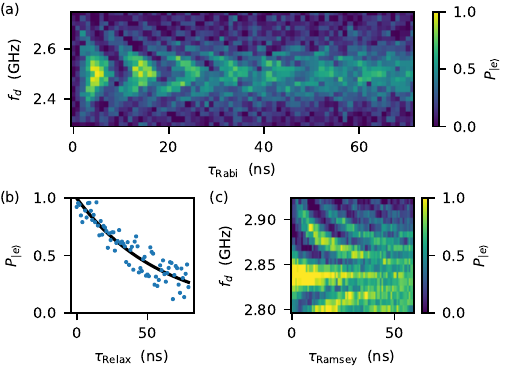}
	\caption{\textbf{Quantum control on device~A.} (a) Rabi oscillations at the gate point $V_g=\SI{-5.246}{V}$. (b) Relaxation measurement at the gate point $V_g=\SI{-6.8108}{V}$ resulting in $T_1=60\pm18\,\text{ns}$. (c) Ramsey chevrons with 8\,ns long square drive pulses at the same gate point as (b), resulting in $T_2^*=36\pm3\,\text{ns}$. These values are about one order of magnitude smaller than achieved for device~B.}
	\label{fig:coherent-control-deviceA}
\end{figure}

\begin{figure}
	\includegraphics[width=\columnwidth]{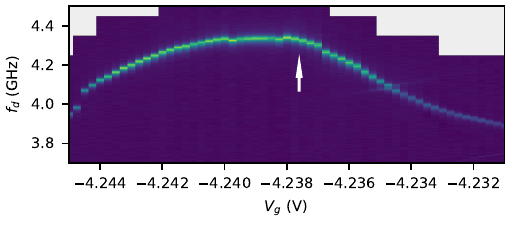}
	\caption{Qubit spectroscopy on device~B of gate region around the gate point $V_g=\SI{-4.2376}{V}$ (white arrow) where data in Fig.~3 in the main text was taken. The color scale is $\arg S_{21}$ after removing the vertical average.}
	\label{fig:coherent-control-2tone}
\end{figure}

\begin{figure}
	\includegraphics[width=\columnwidth]{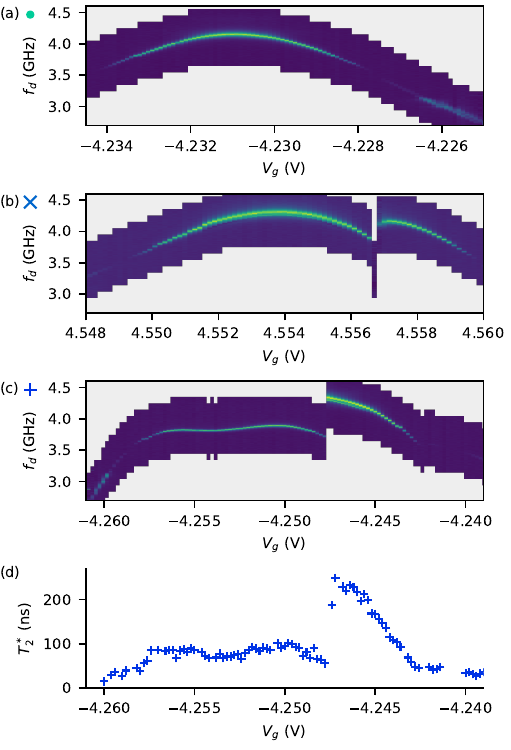}
	\caption{Qubit spectroscopy for gate-dependent coherence time measurements on device~B. (a-c) Qubit spectroscopy corresponding to the data in Fig.~4 in the main text. The color scale is $\arg S_{21}$ after removing the vertical average. (d) Coherence time $T_2^*$ corresponding to (c), showing no increase of $T_2^*$ related to the slope of $f_q(V_g)$. This indicates that coherence time is not gate noise limited. The typical measurement stability is indicated by one gate jump occurring in this data set with a total measurement duration of 54\,hours.}
	\label{fig:gate-dependence-2tone}
\end{figure}

\section{Coherence time measurements without on-chip gate filtering (device A)}
\label{app:coherence-deviceA}

Figure~\ref{fig:coherent-control-deviceA} presents measurements demonstrating coherent control of device~A, with performance representative of the best values obtained across different gate regions. The Rabi chevron in Fig.~\ref{fig:coherent-control-deviceA}a was measured at a gate voltage of \(V_g = \SI{-5.2460}{V}\). The measurements in Fig.~\ref{fig:coherent-control-deviceA}b–c yield a relaxation time of \(T_1 = 60 \pm 18\,\text{ns}\) and a coherence time of \(T_2^* = 36 \pm 3\,\text{ns}\), where uncertainties represent one standard deviation inferred from data scatter. Both measurements were performed at \(V_g = \SI{-6.8108}{V}\).

The qubit frequency in these measurements differs from that in Fig.~\ref{fig:spectroscopy} of the main text, as the data were acquired in separate experimental runs conducted several months apart.

\section{Complementary measurements to Fig.~3 and Fig.~4}
\label{app:spectro-deviceB}

Figure~\ref{fig:coherent-control-2tone} presents qubit spectroscopy of device~B in the gate region surrounding the voltage where the measurements in Fig.~\ref{fig:coherent-control} of the main text were performed.

The relaxation time specified for the measurement displayed in Fig.~3b is obtained by a least-squares fit of the exponential
\begin{equation}
    |S_{21} - S_{21}^{\ket{g}}| = a\, e^{-\tau_\text{Relax} / T_1} + b
\end{equation}
with free parameters $a$, $b$, and $T_1$. The coherence time in Fig.~3d is fitted by the model
\begin{equation}
    |S_{21} - S_{21}^{\ket{g}}| = a\, e^{-\tau_\text{Ramsey}/T_2^*} \cos(2\pi \tau_\text{Ramsey}/T_p + \theta) + b
\end{equation}
with free parameters $a$, $b$, $T_2^*$, the period $T_p$ and phase offset $\theta$.

Figure~\ref{fig:gate-dependence-2tone}a-c shows the qubit spectroscopy corresponding to the gate-dependent coherence data in Fig.~\ref{fig:gate-dependence} of the main text. At each gate voltage, we run an automated pipeline to calibrate qubit readout and excitation. We then extract \(T_1\) from a relaxation measurement and \(T_2^*\) from a Ramsey chevron fit. To increase robustness, the value of $T_2^*$ is the average value over fit results at each drive frequency in the Ramsey chevron. The fit parameters $a$ and $b$ are constrained and the average includes only those fits that meet two criteria: (i) the relative uncertainty is sufficiently small, (ii) the signal-to-noise ratio of the measured data is high, and (iii) the period is short enough to not be strongly correlated with $T_2^*$, i.\,e. $T_p/2<T_2^*$.

The exponential dependence of \(T_2^*\) on qubit frequency, observed in Fig.~\ref{fig:gate-dependence} of the main text, follows approximately
\begin{equation}
T_2^* \approx A \times e^{hf_q/E_C},
\end{equation}
where the proportionality factor $A$ is about \SI{1}{ps}.

\section{Effect of gate noise on \texorpdfstring{$T_2^*$}{T2*}}
\label{app:gate-noise}

One possible limitation of \(T_2^*\) is voltage noise on the top gate, which can induce fluctuations in \(E_J\) and consequently lead to uncontrolled variations in the qubit frequency. Under this hypothesis, dephasing\,---\,and thus the coherence time\,---\,is expected to be more sensitive to gate noise in regions where the qubit frequency strongly depends on \(V_g\), as observed in nanowire gatemons~\cite{Luthi2018}.

However, our measurements (Fig.~\ref{fig:gate-dependence-2tone}d) show that \(T_2^*\) is not correlated with the dependence of \(f_q\) on \(V_g\). Specifically, when the slope \(\mathrm{d}f_q/\mathrm{d}V_g\) vanishes, we do not observe any significant improvement in the coherence time. This strongly suggests that coherence is not limited by gate noise. This conclusion is consistent with the findings presented in the main text, where \(T_2^*\) is attributed to charge noise.

\begin{figure}
	\includegraphics[width=\columnwidth]{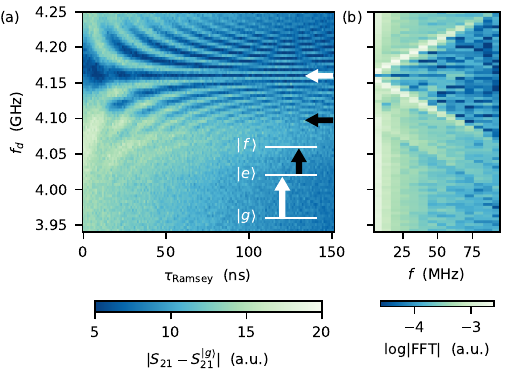}
	\caption{\textbf{Anharmonicity measurement of the qubit on device~B.}
    (a)~Ramsey oscillations after a preceding $\pi$~pulse. The stronger oscillation pattern is at the qubit frequency $f_q=\SI{4.160}{GHz}$ (white arrow). Superposed are oscillations around the transition $f_{ef}=\SI{4.098}{GHz}$ (black arrow) to the second excited state. The $-\alpha=\SI{62}{MHz}$ is the smallest anharmonicity measured on our devices.
    (b)~Fourier transform along the pulse separation $\tau_\text{Ramsey}$. The lines directly correspond to the detuning from the transitions and the detuning associated to the $\ket{e}\rightarrow\ket{f}$ transition is clearly discernible.}
	\label{fig:anharmonicity-ramsey}
\end{figure}

\begin{figure*}
	\includegraphics[width=\textwidth]{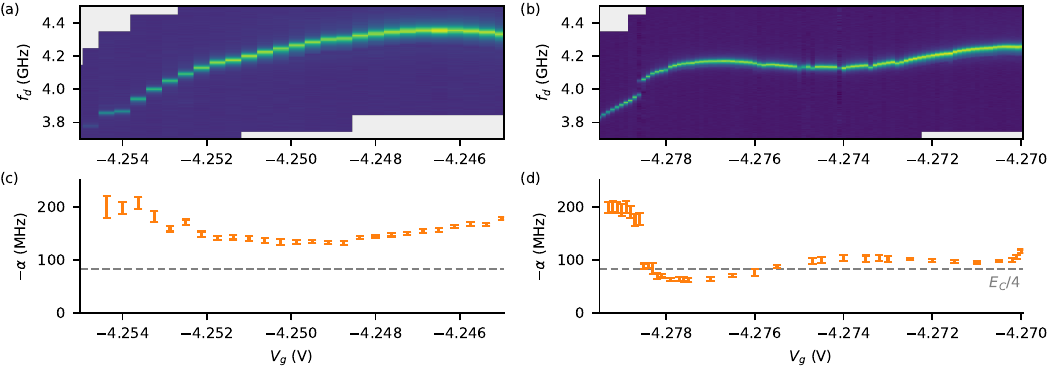}
	\caption{\textbf{Anharmonicity variation of the qubit on device~B.} (a,b)~Qubit spectroscopy in the gate region where the anharmonicity data is taken. The color scale is $\arg S_{21}$ after removing the vertical average. (c,d)~The extracted anharmonicity, which notably is significantly below $E_C=\SI{330}{MHz}$ and for some gate points even below $E_C/4$.}
	\label{fig:anharmonicity}
\end{figure*}

\section{Measurement of qubit anharmonicity using Ramsey interferometry}
\label{app:anharmonicity}

We measure the anharmonicity of device~B using a Ramsey pulse sequence similar to the one described in the main text. The difference lies in an additional \(\pi\) pulse preceding the sequence in order to initialize the qubit in the excited state \(|e\rangle\). When varying the drive frequency $f_d$, the preceding $\pi$ pulse stays at fixed frequency on resonance with the qubit. This measurement, shown in Fig.~\ref{fig:anharmonicity-ramsey}, reveals not only a Ramsey chevron at \( f_{ge} \), corresponding to the \(|g\rangle \rightarrow |e\rangle\) transition, but also at \( f_{ef} \), associated with the \(|e\rangle \rightarrow |f\rangle\) transition. This enables a precise determination of both transition frequencies.

This extraction is only feasible if the charge dispersion of the \(|e\rangle \rightarrow |f\rangle\) transition is on the order of a few MHz or less. This condition is met only for the highest qubit frequencies. Otherwise, excessive dephasing leads to a coherence time that is too short for the second Ramsey oscillations to be resolved.

As discussed in the main text, the anharmonicity of a conventional transmon based on a tunnel junction follows \( -\alpha = E_C \) in the large \( E_J/E_C \) regime. In particular, the bound \( |\alpha| > E_C \) holds for \( E_J/E_C > 20 \)~\cite{Koch2007}, which in our case corresponds to \( f_q > \SI{3.9}{GHz} \) (see Fig.~\ref{fig:gate-dependence} in the main text). Contrary to this, our measurements indicate that the anharmonicity is significantly smaller than the charging energy, \( E_C = \SI{330}{MHz} \), at all measured gate points. We attribute this deviation to a non-sinusoidal current-phase relation, arising from high transmission through the carbon nanotube Josephson junction. Similar behavior has been reported in various other gatemons~\cite{Kringhøj2018,Kringhøj2020,Zheng2024,Sagi2024}.  

The conventional explanation for this effect assumes an ideal short junction, which imposes a lower bound of \( -\alpha \geq E_C/4 \), reached in the case of perfect transmission~\cite{Kringhøj2018}. However, we find an anharmonicity even below \( E_C/4 \) as shown in Fig.~\ref{fig:anharmonicity}. This suggests that a more refined theoretical model, such as the one proposed in Ref.~\cite{Fatemi2024} with weak link of finite length, may be required.  

\section{Temporal stability}
\begin{figure}
    \includegraphics[width=\columnwidth]{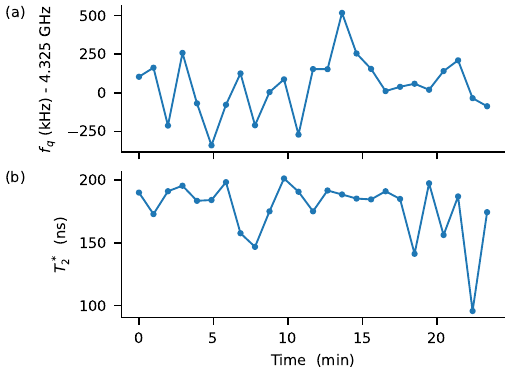}
    \caption{\textbf{Qubit stability in Ramsey measurement.}
    (a)~Variation of qubit frequency over 24\,minutes with an integration time of 1\,minute for each data point. The average value is $f_q=\SI{4.32505}{GHz}$ with a standard deviation of \SI{180}{kHz}.
    (b)~Variation of coherence time in the same measurement. Its average value is \SI{178}{ns} with a standard deviation of \SI{23}{ns}.}
    \label{fig:stability}
\end{figure}

The gatemon's tunability goes hand in hand with the possibility of drifts of $f_q$ and other qubit properties due to changes of $V_g$ or microscopic processes close to the quantum dot Josephson junction~\cite{Feldstein-Bofill2024}.
Gate jumps result in an abrupt change of the qubit frequency as seen in Fig.~\ref{fig:gate-dependence-2tone}c at $V_g=\SI{-4.2477}{V}$ in a measurement of a total duration of 54\,hours. The latter is the typical time scale for a single gate jump in our measurements.

Smaller fluctuations of the qubit properties are explored in Fig.~\ref{fig:stability} showing the variation of $f_q$ and $T_2^*$ in a measurement of Ramsey chevrons with a total duration of 24\,minutes at constant $V_g$. We find a standard deviation of \SI{180}{kHz} for $f_q$ and of \SI{23}{ns} for $T_2^*$ at an integration time of about one minute per point. The scatter reduces with longer integration time, suggesting either a limitation due to the signal-to-noise ratio or fluctuations that are faster than this time scale. This standard deviation of $T_2^*$ is comparable to the uncertainty of $T_2^*$ in Fig.~3d of the main text, extracted for a line cut with a longer duration but not averaged over drive frequencies.



\end{appendix}

\bibliography{references}

\end{document}